# Measuring the Magnetic Flux Density in the CMS Steel Yoke

V. I. Klyukhin

*Skobeltsyn Institute of Nuclear Physics, Lomonosov Moscow State University, RU-119992, Moscow, Russia*

Phone: +41-22-767-6561, fax: +41-22-767-7920, e-mail: Vyacheslav.Klyukhin@cern.ch

N. Amapane

*INFN Torino and the University of Torino, I-10125, Torino, Italy*

A. Ball, B. Curé, A. Gaddi, H. Gerwig, M. Mulders

*CERN, CH-1211, Geneva 23, Switzerland*

A. Hervé, R. Loveless

*University of Wisconsin, WI 53706, Madison, USA*

**Abstract** The Compact Muon Solenoid (CMS) is a general purpose detector, designed to run at the highest luminosity at the CERN Large Hadron Collider (LHC). Its distinctive features include a 4 T superconducting solenoid with 6-m-diameter by 12.5-m-length free bore, enclosed inside a 10000-ton return yoke made of construction steel. The return yoke consists of five dodecagonal three-layered barrel wheels and four end-cap disks at each end comprised of steel blocks up to 620 mm thick, which serve as the absorber plates of the muon detection system. Accurate characterization of the magnetic field everywhere in the CMS detector is required. To measure the field in and around the steel, a system of 22 flux-loops and 82 3-D Hall sensors is installed on the return yoke blocks. Fast discharges of the solenoid (190 s time-constant) made during the CMS magnet surface commissioning test at the solenoid central fields of 2.64, 3.16, 3.68 and 4.01 T were used to induce voltages in the flux-loops. The voltages are measured on-line and integrated off-line to obtain the magnetic flux in the steel yoke close to the muon chambers at full excitations of the solenoid. The 3-D Hall sensors installed on the steel-air interfaces give supplementary information on the components of magnetic field and permit to estimate the remanent field in steel to be added to the magnetic flux density obtained by the voltages integration. A TOSCA 3-D model of the CMS magnet is developed to describe the magnetic field everywhere outside the tracking volume measured with the field-mapping machine. The results of the measurements and calculations are presented, compared and discussed.

# 1 Introduction

The muon system of the Compact Muon Solenoid (CMS) includes a 10,000-ton yoke comprised of the construction steel plates up to 620 mm thick, which return the flux of the 4 T superconducting solenoid and serve as the absorber plates of the muon detection system [1-3].

A three-dimensional (3-D) magnetic field model of the CMS magnet has been developed [4], [5] for the track parameter reconstruction when the detector begins operation. The model is calculated with TOSCA [6] and reproduces within 0.1% [7] the magnetic field measured inside the CMS coil with the field-mapping machine at five central field values of 2, 3, 3.5, 3.8, and 4 T. This model is used to prepare the CMS magnetic field map everywhere outside the measured volume.

A direct measurement of the magnetic flux density in selected regions of the yoke is provided with 22 flux-loops to estimate the uncertainty of utilization of the calculated values for the magnetic field to be used to determine the momenta of muons during the detector operation. The flux-loops of 315÷450 turns have been wound around selected blocks of the CMS yoke plates to measure the magnetic flux changes induced by the "fast" (190 s time-constant) discharges of the CMS coil made possible by the protection system, which is provided to protect the magnet in the event of major faults [8], [9]. An integration technique [10], [11] was developed to reconstruct the average initial magnetic flux density in steel blocks at the full magnet excitations, and the contribution of the eddy currents was calculated with ELECTRA [12] and estimated on the level of a few per cent [13].

In present paper we summarize the results of the magnetic flux measurements done with the flux-loops and compare the obtained results with the calculations performed with the TOSCA CMS magnet model.

# 2 The CMS Magnet Model Description

The CMS magnet model in the configuration used for the magnet surface test is presented in Fig. 1. The model comprises the coil four layers of the superconductor and is separated into two halves of the magnet yoke that consists of five barrel wheels of the 6.995 m inscribed outer radius and 2.536 m width, two nose disks of 2.63 m radius on each side of the coil, three large end-cap disks of the 6.955 m inscribed outer radius on each side of the magnet, and one small end-cap disk of 2.5 m radius on one side of the magnet.

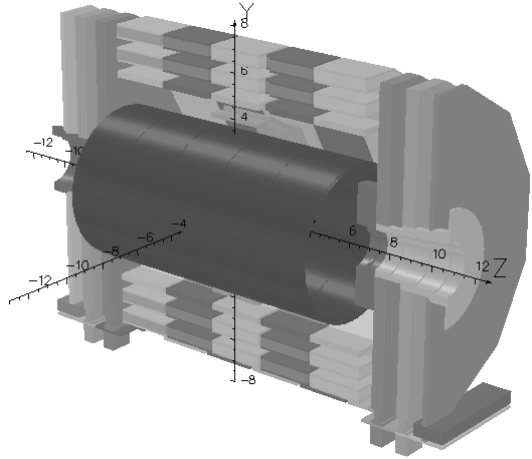

**Fig. 1** 3-D model of the CMS magnetic system (a half of the yoke is shown)

Each barrel wheel except of central one has three layers of steel connected with brackets. The central barrel wheel comprises the fourth most inner layer, tail catcher, made of steel and turned by 5 degrees in the azimuth angle with respect to dodecagonal shape of the barrel wheels. The coordinate system used in the model corresponds to the CMS reference system where the X-axis is directed in horizontal plane toward the LHC center, the Y-axis is upward, and the Z-axis coincides with the superconducting coil axis.

The thickness of the tail catcher blocks started at the inscribed radius of 3.868 m is 0.18 m, the thickness of the first barrel layer is 0.285 m, both second and third barrel layers has a thickness of 0.62 m. The air gap between the tail catcher and the first barrel layer is 0.567 m, the air gap between the first and second barrel layers is 0.45 m, and the air gap between the second and third barrel layers is 0.405 m. All these air gaps are used to install the muon drift tube chambers to register the muon particles.

The barrel wheels are denoted as follows: YB0 is for central wheel, YB±1 are for the next to central wheels, and YB±2 are for two extreme barrel wheels. The air gaps between YB0 and YB±1 are 0.155 m and the air gaps between YB±1 and YB±2 are 0.125 m. Two chimneys for the electrical and cryogenic leads, the barrel wheels feet, and the connecting brackets between the barrel layers are included into the model.

The thickness of two first, YE±1 and YE±2, end-cup disks on each side of the coil is 0.592 m, the thickness of third disks, YE±3, is 0.232 m, and the thickness of forth small disk YE+4 is 0.075 m. The air gaps between YB±2 and YE±1 are 0.649 m, the air gaps between YE±1 and YE±2 are 0.663 m, the air gaps between YE±2 and YE±3 are 0.668 m, and the air gap between YE+3 and YE+4 is 0.664 m. All these air gaps are used to install the muon cathode strip chambers.

Both nose disks, YN±1, are partially inside the coil. The distance between YN–1 and YN+1 is 12.666 m that corresponds to the YE±1 and YN±1 deformation under the magnetic forces at the CMS magnet full excitation. The end-cap disk carts upper plates of 0.1 m thickness and keels are included into the model.

The model comprises 21 conductors, is subdivided into two halves of the CMS yoke and contains in total 7,186,417 nodes of the finite element mesh.

## 2.1 Coil Description

The CMS coil consists of five modules of 2.5 m long, and at the room temperature has the designed length of 12.514 m and inner diameter of 6.3196 m. Four layers of superconductor make the coil thickness of 0.2632 m. In the model the coil dimensions are considered at the temperature of 4°K and the scale factor of 0.99585 is applied to all the dimensions. Changing the shape of the coil under the magnetic pressure is also taken into account. Thus, in the model, the mean radii of the superconductor layers in the central coil module CB0 are 3.18504, 3.25017, 3.3153, and 3.38043 m. In two adjacent coil modules CB±1 the mean radii of the superconducting layers are 2 mm less, and in two extreme coil modules CB±2 the corresponded mean radii are 5 mm less than in the central coil module. There is one missing turn, out of 2180 designed turns in the most inner layer of the CB–2 module between Z=–3.76493 and Z=3.8 m, thus the produced magnetic field is slightly asymmetric with respect to the coil middle plane at Z=0 m.

## 2.2 Steel Magnetic Properties Description

Three different B-H curves of the construction steel of the CMS magnet yoke are used in the model. The first curve describes the magnetic properties of the barrel wheels thick plates in the second and third layers. Second curve describes the magnetic properties of thin plates around the thick plates of the second and third barrel wheels layers, and also the properties of the plates of the first layers and the tail catcher plates of the barrel wheels. Finally, third curve describes the magnetic properties of the nose and end-cap disks, the cart plates and keels.

# 3 Flux-loops description

The CMS magnet yoke is made of construction steel that contains up to 0.17% C, up to 1.22% Mn, and also some Si, Cr, and Cu. To reconstruct the magnetic flux density at the CMS magnet excitations before the fast discharges of the coil, 22 flux-loops have been wound around ten blocks of the bottom 30º azimuthal sectors of the barrel wheels YB0, YB–1, YB–2 and around two bottom 18º azimuthal sectors of thick end-cap disks YE–1 and YE–2. The flux-loops are performed from the flat ribbon cable of 45 wires that has been wound 7÷10 times around the blocks. The areas enclosed by the flux-loops vary from 0.31 to 1.59 m² on the barrel wheels and from 0.5 to 1.13 m² on the end-cap disks. The flux-loops of the barrel wheels are mounted at Z-coordinates of 0, –2.691, –4.224, –5.352, and –6.48 m; the flux loops at the end-cap disks are mounted at the Y-coordinates of –2.8, –4.565, and –6.235 m as shown in Fig. 2.

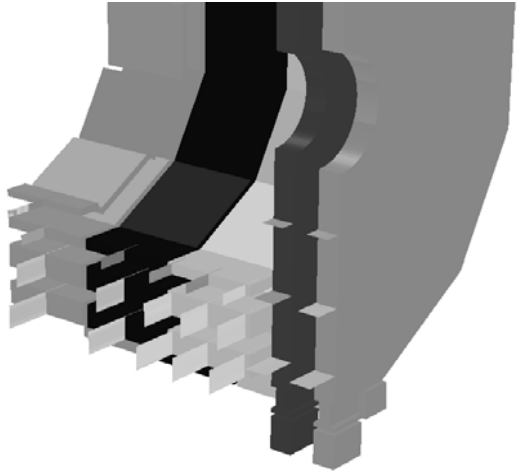

**Fig. 2** Flux-loops locations displayed at X>0 m with its half-areas outside the halves of the YB0, YB–1, YB–2 wheels and YE–1, YE–2 disks

To read out the voltages induced in the flux-loops by changing the magnetic flux in the blocks during the coil fast discharge, seven USB-based DAQ modules USB-1208LS of Measurement Computing [14] with 4 differential 12-bit analog inputs each are used. The USB-1208LS DAQ modules are attached by the USB cables to two network-enabled AnywhereUSB®/5 hubs [15] connected to the personal computer through 3Com® OfficeConnect® Dual Speed Switch 5 [16] sitting on local Ethernet network cable of 90 m.

# 4 Comparison of the Measured and Calculated Magnetic Flux Density in Steel

The average magnetic flux density in steel blocks of the CMS magnet yoke is reconstructed by the off-line integration of the voltages induced in the flux-loops during the coil fast discharges made from the current values of 12.5, 15, 17.55, and 19.14 kA. These current values produce the magnetic flux density in the CMS coil center of 2.64, 3.16, 3.68, and 4.01 T, accordingly. The shapes of the coil current vs. time during the coil fast discharges are displayed in Fig. 3.

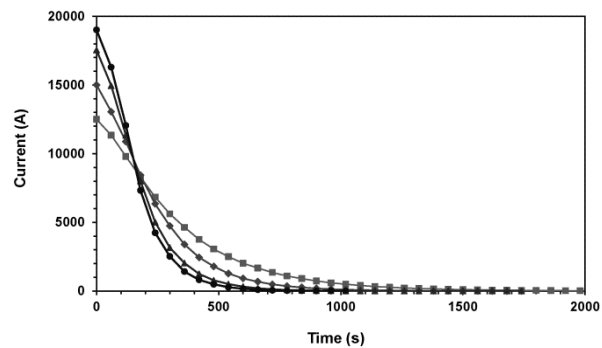

**Fig. 3** CMS coil fast discharges made from 12.5 (squares), 15 (rhombs), 17.55 (triangles), and 19.14 (circles) kA

Those current discharges lead to changing the magnetic flux inside the yoke steel elements that induces the voltages in the flux-loops with the amplitudes up to 3÷4.5 V [7]. The integrations give the magnetic fluxes and then the total averaged field values in steel blocks at the maximum currents being before the coil fast discharges. The precision of the magnetic flux measurements after the off-line voltage integration is 2.1 %. The remanent fields were estimated with the measurements done with the 3-D Hall sensors installed on the steel-air interfaces of the yoke blocks used in the flux-loops measurements. The remanent fields were found to be at the level of a few mT and thus are not added to the magnetic flux densities obtained by the voltages off-line integration.

In Figs. 4–7 the measured values of the magnetic flux density vs. Z- and Y-coordinates are presented and compared with the calculated field values obtained by the magnetic flux density integration within the flux-loops areas in three layers of the barrel wheels and in two end-cap disks for the maximum current values of 17.55 and 19.14 kA. The working current of the CMS detector in present time is 18.164 kA that corresponds to the 3.81 T central field.

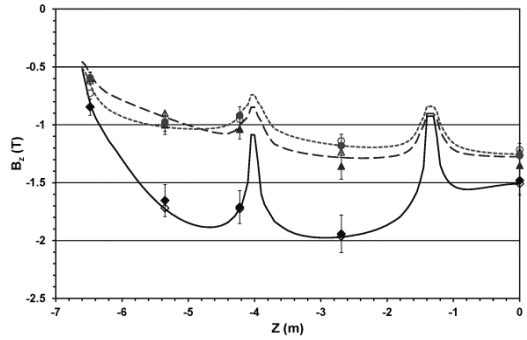

**Fig. 4** Axial magnetic flux density measured (filled markers) and calculated (opened markers) in the first (rhombs), second (triangles), and third (circles) barrel layers vs. Z-coordinate at the current of 17.55 kA. The lines represent the calculated values in a vertical plane X=0 m at the Y-coordinates of –4.7575 (solid line), –5.66 (dashed line), and –6.685 (dotted line) m, corresponded to the middle plains of the first, second, and third barrel layers blocks.

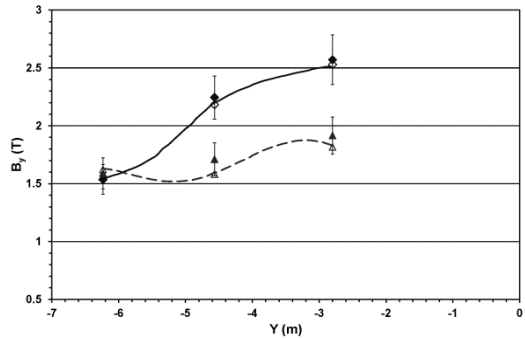

**Fig. 5** Radial magnetic flux density measured (filled markers) and calculated (opened markers) in the first (rhombs), and second (triangles) end-cap disks vs. Y-coordinate at the current of 17.55 kA. The lines represent the calculated values in a vertical plane X=0 m at the Z-coordinates of –7.565 (solid line), and –8.82 (dashed line) m, corresponded to the middle plains of the first and second end-cap disks blocks.

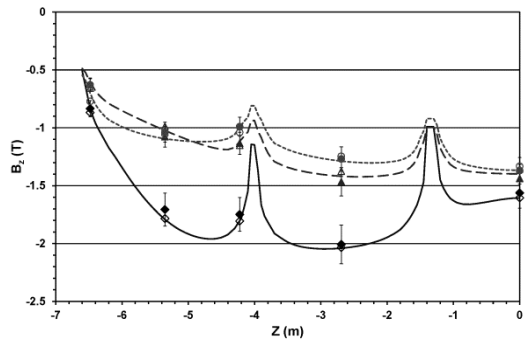

**Fig. 6** Axial magnetic flux density measured (filled markers) and calculated (opened markers) in the first (rhombs), second (triangles), and third (circles) barrel layers vs. Z-coordinate at the current of 19.14 kA. The lines represent the calculated values in a vertical plane X=0 m at the Y-coordinates of –4.7575 (solid line), –5.66 (dashed line), and –6.685 (dotted line) m, corresponded to the middle plains of the first, second, and third barrel layers blocks.

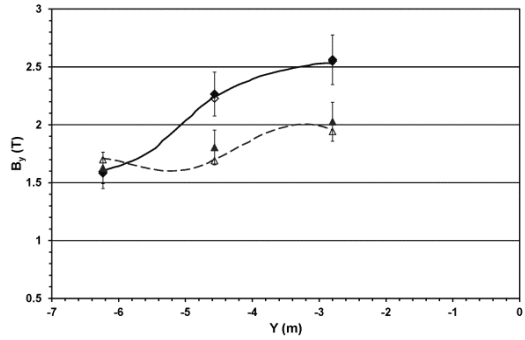

**Fig. 7** Radial magnetic flux density measured (filled markers) and calculated (opened markers) in the first (rhombs), and second (triangles) end-cap disks vs. Y-coordinate at the current of 19.14 kA. The lines represent the calculated values in a vertical plane X=0 m at the Z-coordinates of –7.565 (solid line), and –8.82 (dashed line) m, corresponded to the middle plains of the first and second end-cap disks blocks.

The comparison of the measured and calculated magnetic flux density in steel blocks of the CMS yoke gives the differences between the calculated and measured values as follows: (–0.06±7.48)% in the barrel wheels and (–2.54±3.43)% in the end-cap disks at the maximum current of 17.55 kA; (1.68±7.07)% in the barrel wheels and (–1.21±3.78)% in the end-cap disks at the maximum current of 19.14 kA These differences partially could be described in terms of the eddy currents contributions [13]. The error bars of the measured magnetic flux densities presented at Figs. 4–7 are of ±8.35% and include the errors in the knowledge of the flux-loops geometries and the errors of the measured magnetic fluxes.

## 5 Conclusions

The magnetic flux density is measured in steel blocks of the CMS yoke at the coil central field values of 2.64, 3.16, 3.68, and 4.01 T. The precision of measurements performed with the flux-loops in the CMS yoke steel blocks is a few per cent. The CMS magnet model calculations differ from the field values measured with the flux-loops in steel at the central fields of 3.68 and 4.01 T by (0.07±6.49) % in average that allows using the model to describe the magnetic flux density distribution inside the CMS steel yoke.